\documentclass[acmtog, nonacm]{acmart}

\AtBeginDocument{%
  \fancypagestyle{plain}{%
    \fancyhf{} %
    \fancyfoot[L]{ACM SIGENERGY Energy Informatics Review}%
    \fancyfoot[R]{Volume 6 Issue 2, July 2026}}
  \providecommand\BibTeX{{%
    \normalfont B\kern-0.5em{\scshape i\kern-0.25em b}\kern-0.8em\TeX}}}
    
\let\oldmaketitle\maketitle
\renewcommand{\maketitle}{%
  \oldmaketitle%
  \thispagestyle{plain}%
  \pagestyle{plain}}

\begin{document}

\title{Hot AI in Cold Space: Thermal-Crosstalk-Aware Scheduling for Sustainable Orbital AI Clusters}

\author{Shuyi Chen}
\orcid{0000-0003-0745-4083}
\affiliation{%
    \institution{Southern University of Science and Technology}
    \city{Shenzhen}
    \country{China}
}
\email{chensy8@mail.sustech.edu.cn}

\author{Zhengchang Hua}
\orcid{0000-0002-3970-6129}
\affiliation{
    \institution{Southern University of Science and Technology}
    \city{Shenzhen}
    \country{China}
}
\email{huazc@mail.sustech.edu.cn}

\author{Nikos Tziritas}
\orcid{0000-0002-2091-2037}
\affiliation{
    \institution{University of Thessaly}
    \city{Lamia}
    \country{Greece}
}
\email{nitzirit@uth.gr}

\author{Georgios Theodoropoulos}
\orcid{0000-0002-2091-2037}
\affiliation{
    \institution{Southern University of Science and Technology}
    \city{Shenzhen}
    \country{China}
}
\email{theogeorgios@gmail.com}

\renewcommand{\shortauthors}{Chen et al.}

\begin{abstract}
Terrestrial AI training faces an unsustainable energy and water crisis, positioning Orbital Data Centers (ODCs) as a "zero operational carbon" alternative. However, the sub-$10\mu\text{s}$ communication latency required for synchronized scientific workloads, such as distributed Large Language Model (LLM) training, forces ODCs into extreme physical density, triggering a critical "Proximity-Thermal Paradox." As these high-density systems scale into Monolithic Structures or Proximity Swarms, they suffer from intense thermal-fluid crosstalk (heat traps in shared cooling loops) and thermal-radiative crosstalk (mutual heating that blocks deep-space cooling radiators). If left unmitigated, this persistent heat stagnation not only triggers severe thermal throttling that degrades training throughput, but also induces severe thermal fatigue, drastically shortening hardware lifespans and generating premature space e-waste. To make orbital AI truly sustainable, this position paper challenges traditional uniform load-sharing. We propose the Thermal-Aware Heterogeneity Thesis, which treats spatial cooling variances as a primary resource management dimension. Building on this, we introduce Thermal-Load Balancing (TLB), a software framework that dynamically migrates these intensive workloads to the coolest available units based on instantaneous fluid temperatures or absorbed radiation. Our analysis demonstrates that TLB resolves thermal bottlenecks to restore Model Flops Utilization (MFU), while simultaneously reducing physical thermal stress. Extending the operational lifespan of orbital hardware is crucial to amortize the massive embodied carbon of rocket launches, outlining a necessary pathway to scale orbital AI without accelerating e-waste.
\end{abstract}



\maketitle

\section{Introduction}
Terrestrial AI training faces an unsustainable energy and water crisis, making Orbital Data Centers (ODCs) an attractive "zero operational carbon" alternative \cite{SurveySatelliteComputing_2025, SpaceBasedDataCenter6G_2026}. However, to truly offset the massive \textit{embodied carbon} of rocket launches \cite{DirtyBitsLEOOrbit_2025}, ODCs must maximize their lifetime Model Flops Utilization (MFU) and avoid premature hardware failures. This presents a severe challenge for latency-sensitive, synchronized scientific computations. As a primary example, distributed Large Language Model (LLM) training, particularly when utilizing Tensor Parallelism, demands sub-$10\mu\text{s}$ communication latencies \cite{ScalingIntelligenceDataCenters_2025}. This forces computing nodes into extreme high-density configurations (e.g., 3-meter physical proximity), triggering a critical \textit{Proximity-Thermal Paradox}: the spatial compactness required for low-latency synchronization inevitably leads to intense thermal crosstalk.

This crosstalk is particularly destructive because synchronous distributed workloads rely on "All-or-Nothing" synchronization patterns (e.g., All-Reduce operations in LLM training). Under these strict constraints, a single thermally congested "straggler" node stalls the entire gigawatt-scale cluster \cite{ThermalAwareWorkloadScheduler_2025, DistributedMLLoadBalancingCloud_2020}. If unmitigated, this persistent heat stagnation not only degrades training throughput but also induces severe thermal fatigue, which drastically shortens hardware lifespans and generates premature "space e-waste."

As high-density ODCs scale, they fall into two distinct architectural paradigms, each presenting multi-scale thermal interferences.
\textit{Monolithic Structures} are massive self-assembled units that rely on centralized fluid loops, suffering from \textit{Thermal-Fluid Crosstalk} as heat from high-density GPU tiles accumulates along shared cooling paths and creates downstream heat traps \cite{StarCloudWP_2024}.
Alternatively, \textit{Proximity Swarms} consist of free-flying nodes maintained in tight formations \cite{GoogleSuncatcher_2025}. These swarms suffer from \textit{Thermal-Radiative Crosstalk}, where geometric shadowing and mutual heating block deep-space cooling windows, particularly for nodes trapped in the cluster's core \cite{ThermalDesignSatellitePayload_2025}.

\begin{figure}
    \centering
    \includegraphics[width=0.99\linewidth]{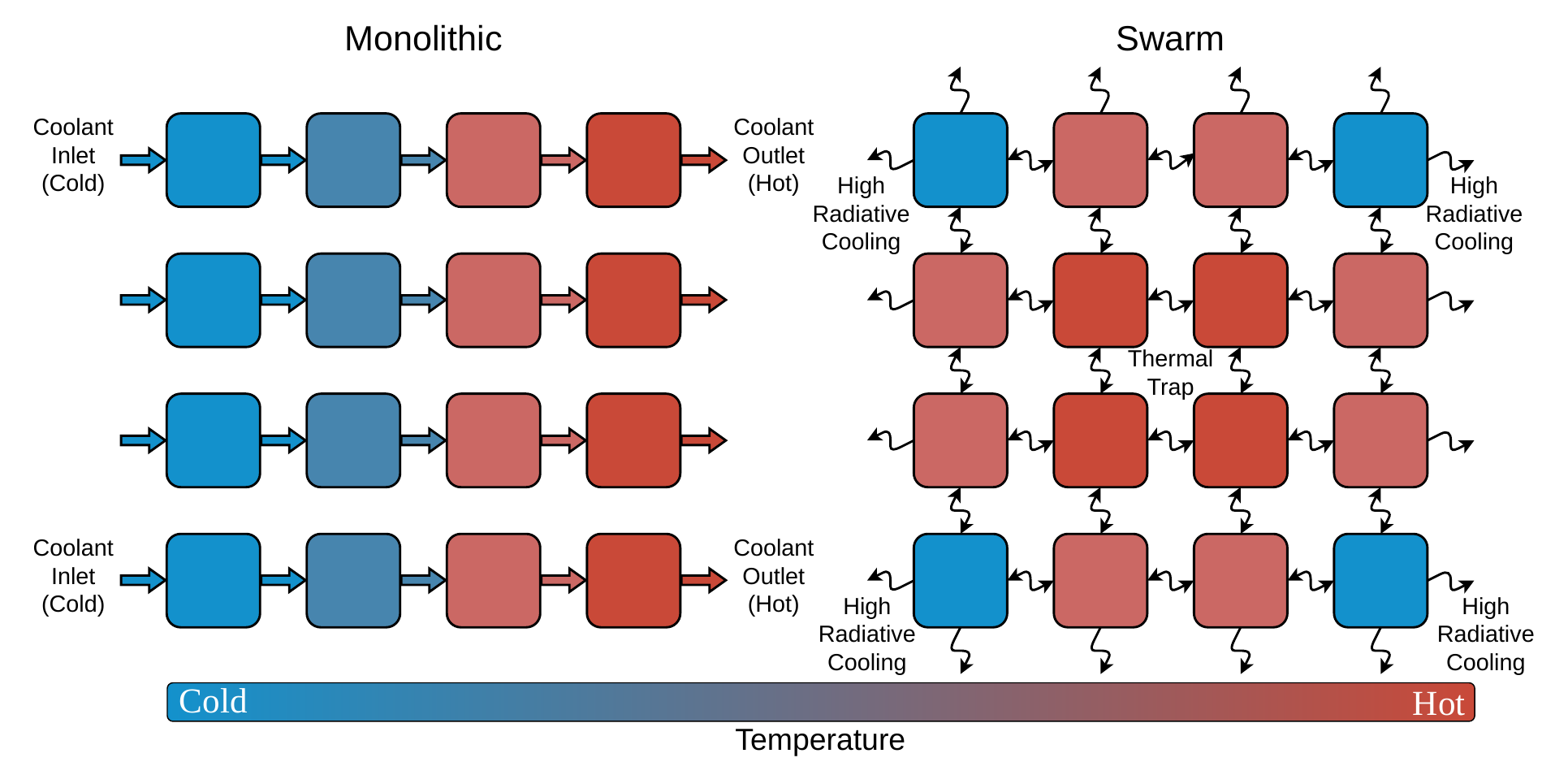}
    \caption{Two ODC architectural paradigms: Monolithic Structures with centralized cooling loops and Proximity Swarms of free-flying nodes.}
    \Description{A diagram illustrating two ODC architectures. On the left, a monolithic structure shows centralized cooling loops with heat exchangers. On the right, a proximity swarm shows multiple free-flying satellite nodes communicating and radiating heat in a tight formation.}
    \label{fig:sdc_arch}
\end{figure}

In both paradigms, traditional load-balancing frameworks force structural and operational uniformity, which inevitably leads to synchronous bottlenecks. To make orbital AI physically and environmentally sustainable, we introduce the \textit{Thermal-Aware Heterogeneity Thesis}. We argue that ODCs must reject uniform load-sharing and instead treat spatial and temporal cooling variance as a primary, schedulable computational resource.

Building on this, we propose the Thermal-Load Balancing (TLB) paradigm. TLB is a software orchestration framework that dynamically migrates computation blocks to the coolest available units based on instantaneous fluid temperatures or absorbed radiation. By proactively shifting tasks, TLB actively resolves heat traps, prevents the emergence of thermal stragglers, and mitigates cyclical thermal stress on the silicon.

This paper outlines a pathway to scale orbital AI without accelerating space e-waste. Specifically, we make the following contributions:
\begin{itemize}
    \item We define the \textit{Proximity-Thermal Paradox} in ODCs, detailing how extreme density induces severe Thermal-Fluid and Thermal-Radiative Crosstalk.
    \item We introduce the \textit{Thermal-Aware Heterogeneity} thesis, establishing that exploiting cooling variance is essential to preventing thermal fatigue and straggler-induced MFU degradation.
    \item We propose the \textit{Thermal-Load Balancing} framework, providing workload migration strategies to resolve heat traps in both monolithic flow paths and swarm radiative fields.
    \item We formulate an evaluation methodology, modeling a dense multi-node mesh, to demonstrate that TLB restores MFU and extends hardware lifespans by reducing cyclic thermal stress, thereby amortizing the embodied carbon of ODCs.
\end{itemize}

\section{Related Work}
The development of the TLB paradigm intersects several established research domains. To contextualize the novelty of our approach, this section reviews the state-of-the-art in space-based distributed computing, spacecraft thermal control mechanisms, and terrestrial AI load-balancing strategies.

\subsection{Distributed Computing in Orbital Data Centers}
The paradigm of satellite operations is rapidly shifting from simple data collection and downlink to on-orbit data processing. To support latency-sensitive and compute-intensive applications such as distributed machine learning and autonomous navigation, researchers have proposed the concept of ODCs \cite{SurveySatelliteComputing_2025, IntegrationDataCenterSDCN_2022, SpaceBasedDataCenter6G_2026}.
Recent studies have explored various architectures for space-based edge computing, including hierarchical multi-orbit systems, micro datacenters using torus topologies, and satellite-terrestrial integrated networks \cite{SatelliteTerrestrialEdge_2020, HighPerfOnOrbitComputing_2025, MicroDatacenterSpace_2025}.
Specific to application workloads, frameworks for on-orbit remote sensing and AI task offloading have been proposed to minimize delay and energy consumption \cite{SatelliteEdgeComputing_2025, OnOrbitTaskOffloading_2023, LEOCloudAIComputing_2023}. By clustering multiple satellite nodes equipped with commercial-off-the-shelf accelerators, ODCs can circumvent the bandwidth limitations and high latency of terrestrial downlinks.

However, supporting synchronous workloads like distributed LLM training introduces a distinct networking challenge. Terrestrial AI clusters rely on high-speed backplanes or Optical Circuit Switches \cite{JupiterEvolving_2022} to achieve the sub-$10\mu\text{s}$ latencies required for synchronous operations like Tensor Parallelism.
Replicating this in space via Free-Space Optical (FSO) links forces ODCs into extremely tight spatial topologies \cite{OpticalWirelessComm_2026}. Current space-based computing literature primarily focuses on asynchronous tasks (e.g., remote sensing offloading) \cite{OnOrbitTaskOffloading_2023}, largely overlooking the severe thermal consequences of deploying these tightly coupled, synchronous networks.

\subsection{Spacecraft Thermal Management}
Thermal management in the vacuum of space relies predominantly on radiative heat transfer, supplemented by internal conductive and fluid loop systems, as well as phase change materials and passive thermal coatings \cite{ReviewThermalManagementSpacecraft_2024, SpacecraftThermalControl_2021, PassiveThermalControl_2023, PhaseChangeMaterialsSpacecraft_2013}.
Traditional spacecraft thermal design assumes a relatively unobstructed view factor to deep space and isolated internal heat generation. Current literature primarily addresses thermal control for isolated satellites or loosely formed constellations. While terrestrial HPC studies have explored heat accumulation in pumped liquid loops \cite{LiquidCoolingDataCenters_2022}, and aerospace studies have modeled radiative mutual heating in simple formations \cite{ThermalDesignCubesats_2021}, existing space thermal frameworks do not account for the multi-scale, dynamic thermal congestion inherent to GW-scale monoliths or densely packed swarms.
Hardware over-provisioning adds prohibitive mass, severely exacerbating the embodied carbon footprint of the mission \cite{DirtyBitsLEOOrbit_2025}, and mechanical orbital reconfiguration consumes limited propellant.
Current space thermal control literature rarely addresses the rapid, cyclic thermal fatigue induced by dynamic computational workloads, which accelerates hardware degradation and generates space e-waste. A software-defined approach to thermal management is highly desirable but remains underexplored.

\subsection{Conventional Load Balancing in Distributed AI}
In terrestrial data centers, distributed scientific workloads such as AI training via data or pipeline parallelism rely on load-balancing frameworks that optimize for computational throughput, memory capacity, and network bandwidth \cite{StaticLoadBalancingDNN_2020, DynamicLoadBalanceDNN_2022, DistributedMLLoadBalancingCloud_2020}.
Recent advancements in terrestrial networks also employ intelligent, machine learning-based dynamic load balancing to manage heterogeneous resources and reduce energy expenditure in complex ecosystems like cloud services and edge networks \cite{SurveyMLLoadBalancing_2022, EnergyEfficientAILoadBalancing_2022, AdvancedTechniquesAIDistribution_2024, DistributedAssignmentDNNEdge_2022, OptimizingDemandResponseEV_2024}.
While a distinct body of literature addresses thermal-aware task placement in terrestrial cloud data centers to minimize cooling energy (i.e., operational carbon) \cite{ThermalModelVMPlacement_2024, ThermalAwareVMScheduling_2023}, these models are heavily anchored in convective air-cooling dynamics (e.g., hot-aisle containment) and active HVAC pump logic. They cannot be translated to ODCs, which rely purely on highly variable radiative heat rejection and closed-loop monolithic flow paths.
Since ODCs are powered by solar arrays (effectively zero operational carbon), load balancing must pivot from optimizing electricity expenditure to minimizing thermal stress and maximizing hardware lifespan to amortize embodied carbon.

When applied to ODCs, conventional load balancers distribute tasks symmetrically or based purely on static hardware specifications, completely ignoring the dynamic and spatially heterogeneous thermal environment of the satellite cluster. As a result, nodes with poor cooling margins quickly hit hard temperature thresholds ($T_{hard}$), triggering aggressive hardware thermal throttling. Because synchronized scientific training requires strict barrier synchronization between nodes (e.g., via All-Reduce operations), the thermal throttling of a single central node cascades, delaying the entire global step, a phenomenon often referred to as the straggler problem \cite{DistributedMLLoadBalancingCloud_2020, TamingUnbalancedTraining_2020}. This highlights the critical need for an orbital-specific thermal-aware scheduling paradigm.

While the vision of ODCs is rapidly maturing, the intersection of spatial topology, thermal dissipation, and distributed AI workload scheduling remains a critical blind spot. Existing literature addresses these challenges in isolation: thermal engineers focus on hardware radiators, network architects optimize for link latency, and AI researchers design load balancers for terrestrial facilities with homogeneous cooling. Our work bridges this gap by proposing that thermal heterogeneity is not just a hardware constraint, but a primary software-schedulable resource.

\section{System model}
\subsection{Orbital Data Center Model}
We model ODCs as high-density satellite clusters and consider two distinct architectural paradigms.
Let $\mathcal{N} = \{1, \dots, N\}$ be the set of satellite nodes. Each node $i$ is characterized by its thermal capacitance $C_i$, dynamic power limits ($P_{idle}$ and peak $P_{max}$), and maximum processing speed $S_{max}$. Heat dissipation properties differ significantly depending on the architecture:

\textbf{Monolithic Structures}. The monolithic ODC features a rigid, large-scale (e.g., MW-to-GW scale) macro-structure shaped like a massive unified container, similar to conceptual tether-based orbital space stations (Starcloud-4) \cite{StarCloudWP_2024}. Power generation and heat dissipation for the entire ODC are managed globally by the primary structure via externally mounted large-area heat radiators and solar arrays. Standalone compute units are housed within this structure and managed by a centralized liquid cooling system.

As shown in the left part of Figure~\ref{fig:sdc_arch}, multiple coolant pipes are routed through the data center, pumping coolant through the compute units before returning it to the heat exchanger. Along the pathway of each coolant loop, upstream nodes receive cold coolant, while downstream nodes receive pre-heated coolant, creating an inherent thermal imbalance across the compute units.

\textbf{Proximity Swarms}. The proximity swarm ODC architecture consists of multiple independent satellite computing nodes flying in a tight formation, similar to Google's Project Suncatcher concept \cite{GoogleSuncatcher_2025}. These nodes form a dense constellation with inter-node distances typically ranging from hundreds of meters to a few kilometers, communicating via high-bandwidth FSO links, as shown in the right part of Figure~\ref{fig:sdc_arch}.

Each satellite in the swarm is equipped with its own solar panels and heat radiators. Although each node utilizes independent passive radiative cooling, the nodes within the constellation are physically close enough to thermally influence one another. Due to view factor occlusion caused by neighboring nodes, their radiators experience varying effective radiating areas. Warmer nodes may also radiate waste heat directly toward their neighbors. Together, these factors establish a dynamic thermal imbalance across the distributed swarm.

\subsection{Heat Dissipation and Thermal Crosstalk Models}
Let $P_{in}$ and $T_{instant}$ denote the instantaneous chip power and temperature. We link computational load to heat generation non-linearly:

\begin{equation}
    P_{in} = P_{idle} + (P_{max}-P_{idle}) \left( \frac{S_{instant}}{S_{max}} \right)^{\gamma}
\end{equation}

The transient thermal state is dictated by the net heat flow $P_{in} - P_{out}$, where heat dissipation $P_{out}$ is architecturally dependent:

\textbf{Thermal-Fluid Crosstalk (Monolithic)}. Downstream nodes are cooled by fluid pre-heated by upstream components:
\begin{equation}
    T_{fluid}^{(i)} = T_{fluid}^{(i-1)} + \eta P_{out}^{(i-1)}
\end{equation}
This degrades the downstream convective heat dissipation rate, creating downstream heat traps:
\begin{equation}
    P_{out}^{(i)} \propto T_{instant}^{(i)} - T_{fluid}^{(i)}
\end{equation}

\textbf{Thermal-Radiative Crosstalk (Swarms)}. Dense formations rely on deep-space radiation governed by the Stefan-Boltzmann law:
\begin{equation}
    P_{out} \propto T_{instant}^4 - T_{amb}^4
\end{equation}
Mutual geometric shadowing severely restricts this, which we model via an Effective View Factor $\rho_i \in (0, 1]$. Peripheral nodes retain larger effective radiating areas (higher $\rho_i$), while core nodes suffer severe occlusion (significantly lower $\rho_i$), triggering immediate heat traps.

Note that our current formulation employs a time-averaged, steady-state deep space environment abstraction. We deliberately isolate the endogenous thermal crosstalk (heat generated strictly by the AI payload and throttled by topological proximity) from exogenous orbital thermodynamics. Highly transient space environment factors, including direct solar flux, Earth infrared emissions, albedo effects, orbital eclipse cycles, and complex 3D spacecraft attitude geometries, are thus abstracted out of this proof-of-concept simulation.
This abstraction is an intentional, conservative baseline. Exogenous variables introduce severe high-frequency thermal transients that will inevitably exacerbate the thermal variance across the cluster. They do not negate the fundamental mechanics of the Proximity-Thermal Paradox; rather, they amplify the necessity of software-level thermal orchestration. Integrating high-fidelity, dynamic orbital mechanics and 3D ray-tracing into the thermal-compute co-simulator represents the immediate trajectory for future work.

\subsection{Throttling and Lifespan Degradation Models}
To avoid hardware damage, nodes enforce a soft temperature threshold ($T_{soft}$) and a hard threshold ($T_{hard}$). When $T_{instant}$ exceeds $T_{soft}$, the processing speed $S_{instant}$ scales down linearly from $S_{max}$, capping at a minimum safe speed $S_{min}$ once $T_{hard}$ is reached.

We also model thermal-induced hardware degradation using the standard Arrhenius equation, where the Mean Time To Failure (MTTF) is exponentially dependent on temperature:
\begin{equation}
    MTTF \propto \exp\left(\frac{E_a}{k_B T}\right)
\end{equation}
Here, $E_a$ is the activation energy (e.g., 0.685 eV) and $k_B$ is the Boltzmann constant. Because of this exponential relationship, even minor reductions in peak operating temperatures yield disproportionately large extensions in hardware lifespan.
\subsection{Workload Model with Asymmetric Data Parallelism}
To accelerate training when memory is sufficient, we apply Data Parallelism. Assuming every computing node holds a full replica of the model, the training data batch is split across different nodes. Each node processes a certain volume of data, determined by the configured micro-batch size.

For distributed workloads executing via Data Parallelism, the global batch (size $B_{global}$) is partitioned such that each node processes a micro-batch of $b_i$ tokens. The computation time is directly proportional to $b_i$ and inversely proportional to the thermally throttled speed $S_i(T)$.

A global training step strictly requires gradient synchronization (e.g., All-Reduce). This means the total step latency is dictated entirely by the single slowest node in the cluster:

\begin{equation}
    t_{step} = \max_i \left( t_{comp}(i) + t_{comm} \right)
\end{equation}

This mathematical “max” operation establishes the severe consequence of the Proximity-Thermal Paradox, where a single node suffering from intense thermal crosstalk will experience degraded $S_i(T)$, thereby stalling the entire ODC cluster and causing a global collapse in MFU.

While the physical density of ODCs is driven by the strict latency bounds of Tensor Parallelism, simulating microsecond-level optical routing alongside macroscopic thermodynamics is computationally intractable. Therefore, we utilize Data Parallelism as a macro-scale surrogate model; since both rely on strict barrier synchronization, the straggler penalty induced by thermal throttling is structurally isomorphic.

\section{The Thermal-Load Balancing Framework}

\subsection{A Generalized Orchestration Architecture}
TLB is designed as a generalized, closed-loop orchestration framework, decoupling the load-balancing \textit{mechanism} from any specific scheduling \textit{policy}. It continuously operates through a control loop of thermal telemetry, capability profiling, and asymmetric workload slicing. TLB aggregates real-time physical metrics like instantaneous silicon temperatures, local fluid pre-heating levels, and dynamic view factor occlusions. These raw thermal data are translated into an abstract capability score ($w_i$) that quantifies each node's real-time thermal margin. Finally, TLB dynamically adjusts the distribution of the AI workload (e.g., micro-batch sizes in Data Parallelism) to match the profiled capabilities.

This abstraction allows ODC operators to plug in arbitrarily complex decision engines, ranging from convex optimization solvers to Deep Reinforcement Learning agents, without altering the underlying distributed AI communication backends.

\subsection{Proof-of-Concept: Proportional Heuristic}
For this position paper, rather than deploying a computationally expensive global optimization solver, we implement a lightweight proportional heuristic. Standard Data Parallel frameworks enforce a uniform batch distribution, allocating $B_{global}/|\mathcal{N}|$ tokens to each node. This severely penalizes the entire cluster when nodes with degraded cooling drop their processing speed $S_i(T)$. TLB breaks this uniformity by calculating the thermal capability score $w_i$. The workload $b_i$ assigned to node $i$ is strictly proportional to its score:
\begin{equation}
    \label{eq:thermal_capability_score}
    b_i = 1 + \left\lfloor (B_{global} - |\mathcal{N}|) \cdot \frac{w_i}{\sum_{k \in \mathcal{N}} w_k} \right\rfloor
\end{equation}
where every node receives at least one baseline unit of work to maintain gradient synchronization participation, and any fractional rounding remainders are allocated to the highest-scoring nodes.

We define an aggressiveness multiplier $\alpha$ and compute the capability scores based on the architectural paradigm:

\textbf{Monolithic Flow-Path Priority}. We prioritize upstream nodes that receive the coldest fluid. Let $L_{pipe}$ be the total number of nodes on a cooling pipe, and $idx_i$ be the zero-based index of node $i$ along that pipe. The capability score is:
\begin{equation}
    \label{eq:flow_score}
    w_i = 1.0 + \alpha \cdot (L_{pipe} - idx_i)
\end{equation}

\textbf{Swarm Radiative Priority}. We prioritize nodes on the periphery of the cluster with an unobstructed view of deep space. By directly utilizing the Effective View Factor ($\rho_i$) introduced in our thermal model, which inherently accounts for geometric shadowing, the capability score is:

\begin{equation}
    \label{eq:radiative_score}
    w_i = 1.0 + \alpha \cdot \rho_i
\end{equation}

By proactively shifting the heaviest computational burdens to nodes with high $w_i$, TLB inherently limits the heat generation of central or downstream nodes, preventing them from hitting $T_{soft}$ and triggering the straggler effect.

We explicitly state that this proportional heuristic is a \textit{baseline feasibility demonstration}, not a terminal algorithmic solution. While the heuristics defined in Equations~\ref{eq:thermal_capability_score}, ~\ref{eq:flow_score}, and ~\ref{eq:radiative_score} do not provide strict mathematical guarantees for boundary safety under highly non-linear thermal transients, they serve as a necessary diagnostic probe. Their purpose is to empirically validate the solvability of the Proximity-Thermal Paradox via asymmetric workload slicing, without obscuring the fundamental system dynamics behind a black-box solver.

\subsection{Integration with the Distributed AI Stack}
To transition TLB from a theoretical model to a deployable system, it must bridge the gap between hardware telemetry and the AI application layer. At the hardware level, TLB leverages standard space-grade baseboard management controllers to poll instantaneous silicon temperatures and coolant states. At the application layer, TLB interfaces with distributed AI frameworks (e.g., PyTorch DistributedDataParallel or Megatron-LM). Instead of relying on a static \texttt{DataLoader} that evenly shards the dataset, TLB implements a \textit{Thermal-Aware Data Sampler}. When the profiling engine updates the capability scores $w_i$, the sampler dynamically adjusts the micro-batch size $b_i$ for the upcoming training steps.

Dynamic workload slicing introduces overheads from altering micro-batch sizes, such as memory pre-allocation and graph recompilation penalties. While hardware throttling might appear as a zero-overhead alternative terrestrially, it induces severe cyclical thermal fatigue in ODCs. This thermal fatigue drastically shortens hardware lifespan, undercutting the launch's massive embodied carbon investment.

To ensure that the overhead of dynamic batching does not cannibalize these critical lifespan gains, TLB relies on emerging dynamic-shape compilers (e.g., PyTorch 2.0 \texttt{torch.compile}) or pre-compiled computational graph buckets corresponding to a discrete set of allowable batch sizes. This guarantees that thermal-induced workload shifts are executed with near-zero software overhead, making asymmetric scheduling strictly beneficial.

Executing the TLB control loop at every training step would introduce unacceptable synchronization delays. Therefore, TLB employs a hybrid dual-loop policy. First, workload redistribution occurs at coarse-grained intervals (e.g., epoch boundaries) to handle slow-moving thermal dynamics like orbital eclipses. Second, because the static proportional heuristic lacks mathematical guarantees to prevent over-allocation in unpredictable thermal states, TLB deploys a rigid, event-driven safety net. This net operates asynchronously from the training loop. If hardware telemetry detects a node rapidly approaching $T_{soft}$, the safety net forcibly triggers an immediate asynchronous task evacuation, acting as an absolute physical backstop. This completely prevents catastrophic threshold violations, ensuring that the absence of algorithmic optimality never compromises physical hardware safety.

\section{Performance Evaluation}
To validate the Thermal-Aware Heterogeneity thesis, we implement a proof-of-concept simulation of the TLB framework. Since this paper focuses on the necessity of a paradigm shift rather than proposing an optimal algorithmic solver, we benchmark a proportional variation of TLB against a standard homogeneous load balancer.

\subsection{Simulation Methodology}
To model the multi-scale thermal dynamics of ODCs, we developed a time-stepped thermal-compute co-simulator\footnote{\url{https://github.com/Sukiiichan/sdc-sim}} that evaluates distributed scientific workloads executing via Data Parallelism.

For the \textbf{Monolithic} architecture, 64 nodes are distributed across 8 liquid cooling pipes (8 nodes daisy-chained per pipe). Downstream nodes process coolant that has been pre-heated by upstream nodes.
For the \textbf{Proximity Swarm} architecture, 36 independent satellites fly in a dense $6\times6$ planar grid formation. We implement a radiative view factor model where nodes dynamically cast thermal shadows on each other; edge nodes retain larger effective radiating areas (higher $\rho_i$), while the core nodes suffer severe radiative occlusion (significantly lower $\rho_i$).

We compare two scheduling paradigms: \textbf{Baseline (Uniform)} shards the global micro-batch evenly across all nodes, mimicking standard PyTorch DDP behavior; \textbf{TLB (Proportional)} implements the proportional allocation heuristic. We track instantaneous hardware temperatures, dynamic processing speeds (throttling), and global step synchronization latency.

\subsection{Mitigating Heat Traps (Spatial Analysis)}
\begin{figure}
    \centering
    \includegraphics[width=0.99\linewidth]{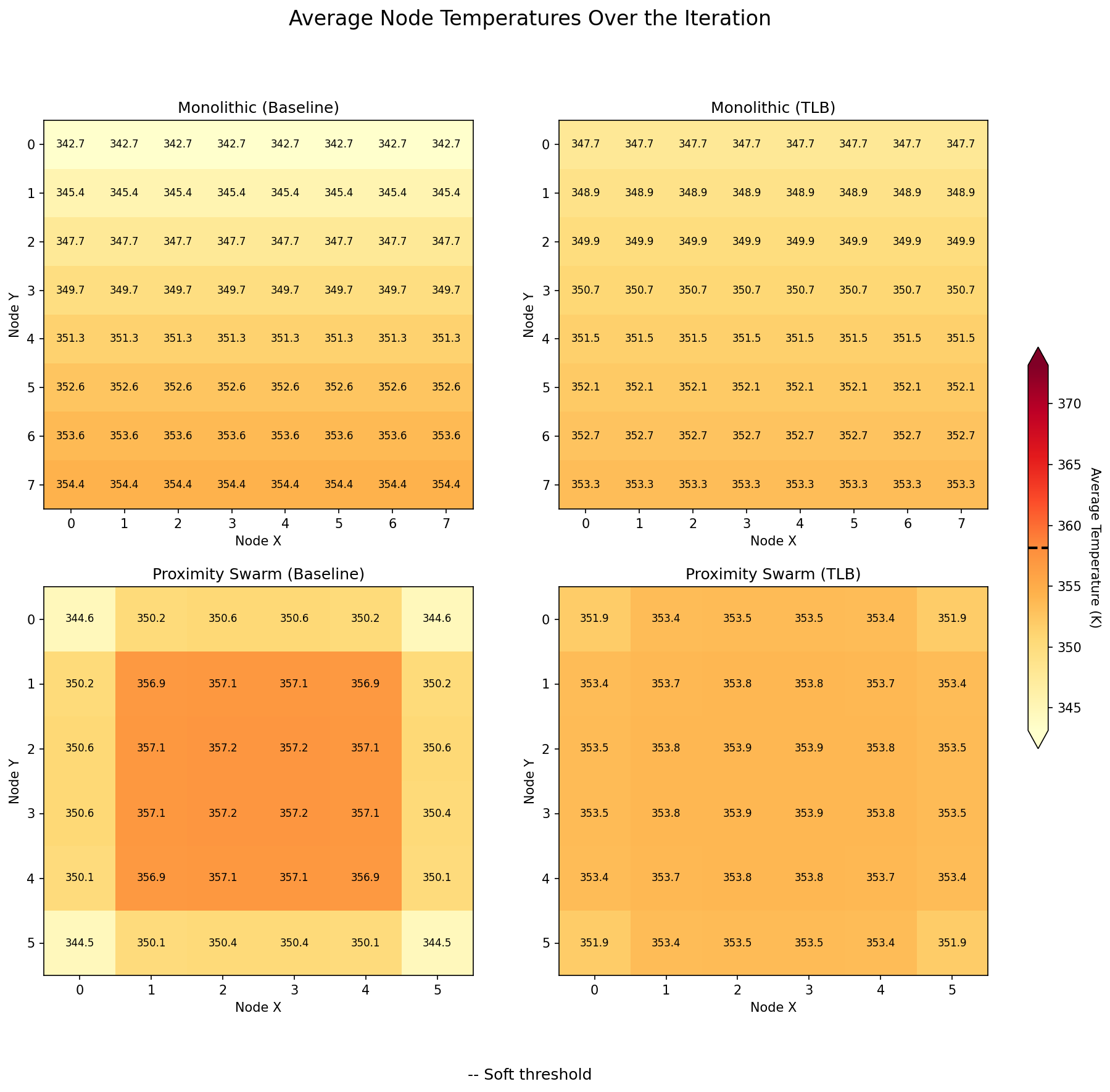}
    \caption{Average Node Temperatures Over the Iteration (Baseline vs. TLB)}
    \Description{Two sets of heatmaps comparing node temperatures for Monolithic and Swarm architectures under Baseline and TLB scheduling. The Baseline heatmaps show severe thermal hotspots, while the TLB heatmaps show significantly more uniform temperature distributions, eliminating the hotspots.}
    \label{fig:heatmap}
\end{figure}

The spatial thermal distribution reveals the fundamental flaw of homogeneous scheduling in ODCs. Figure~\ref{fig:heatmap} shows the average node chip temperatures in the Monolithic and Proximity Swarm architectures. Under the Baseline policy, the \textit{Proximity-Thermal Paradox} manifests rapidly. In the Monolithic configuration, heat progressively accumulates along the cooling loops, pushing the downstream nodes (Row 7) into thermal saturation (up to 354.4 K / 81.3$^\circ$C). In the Swarm configuration, the core nodes become severe heat traps reaching 357.2 K (84.1$^\circ$C) due to high radiative view factor occlusion, while the periphery satellites remain significantly cooler (344.5 K / 71.4$^\circ$C).

TLB dynamically maps the thermal topology and deliberately shifts the heaviest computational burdens to the upstream Monolithic nodes and the periphery Swarm nodes. This asymmetric workload slicing neutralizes the cooling variance. For the Monolithic setup, the maximum temperature drops to 353.3 K (80.2$^\circ$C). For the Swarm, the extreme thermal gradient is flattened, with core temperatures reducing to 353.9 K (80.8$^\circ$C) and edge temperatures rising to 351.9 K (78.8$^\circ$C), effectively eliminating spatial heat traps.

\subsection{Prioritizing Hardware Survivability Over Throughput}

\begin{figure}
    \centering
    \includegraphics[width=0.99\linewidth]{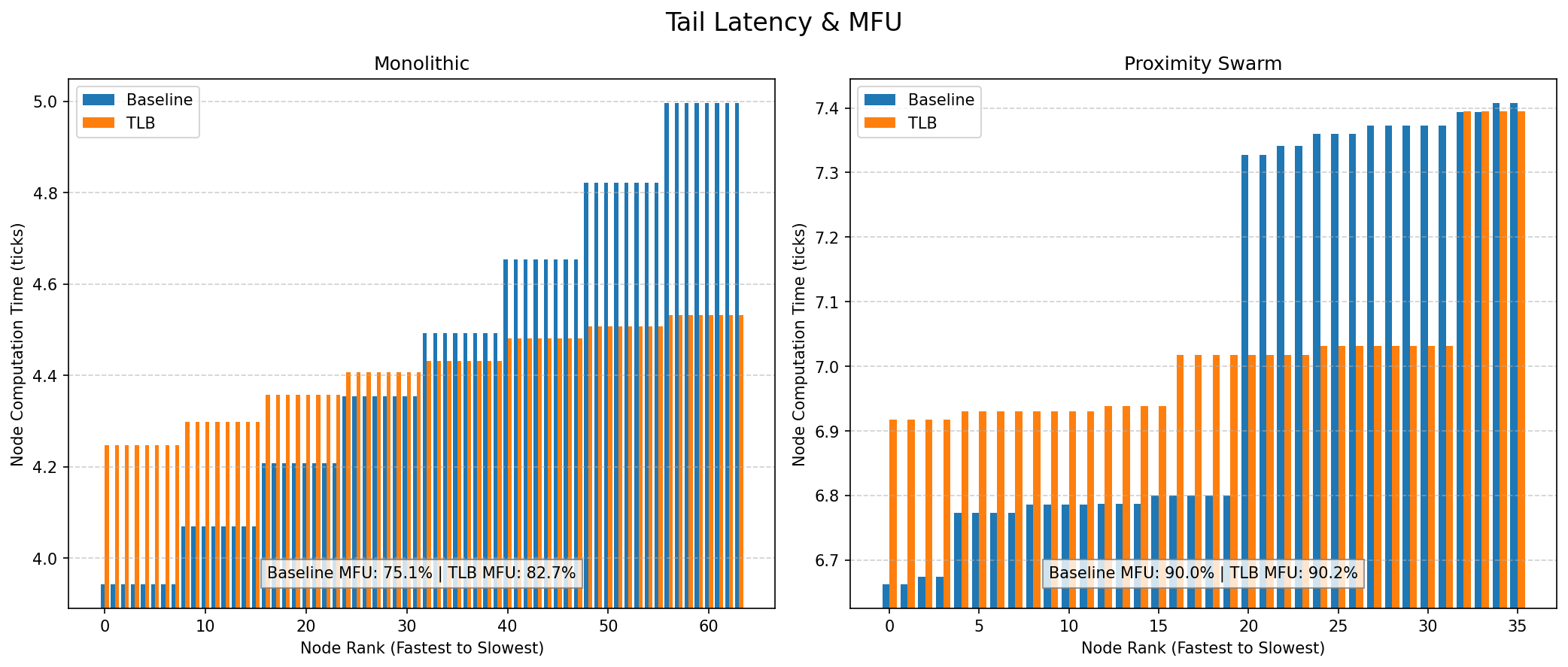}
    \caption{Tail Latency \& MFU of node computation times (Baseline vs. TLB).}
    \Description{Bar charts showing tail latency and MFU for Baseline and TLB strategies. TLB demonstrates reduced tail latency and higher MFU compared to the Baseline, particularly highlighting the flattened computation time across nodes in the swarm architecture.}
    \label{fig:tail_latency}
\end{figure}

The resolution of spatial heat traps shifts the Return on Investment metric for ODCs. Terrestrial data center optimizations traditionally prioritize instantaneous computational throughput (e.g., MFU). However, in the zero-operational-carbon environment of solar-powered orbital swarms, energy is abundant, but hardware lifespan is strictly finite. Thus, a primary metric of success is \textit{hardware survivability} to amortize the massive embodied carbon of the rocket launch.

\textbf{Extending Lifespan to Amortize Embodied Carbon}. Under the uniform Baseline policy, central nodes experience severe thermal fatigue due to relentless temperature oscillation and sustained peak temperatures (e.g., 357.2 K). This cyclical thermal stress rapidly degrades silicon dies and solder joints, accelerating premature space e-waste. By smoothing out thermal gradients and capping peak temperatures at safe operational limits, TLB prevents these nodes from entering cyclical throttling regimes.

Using an apparent activation energy ($E_a$) of 0.685 eV, our Arrhenius-based degradation analysis demonstrates that TLB directly extends the MTTF of the most thermally constrained nodes. We observe a 6.15\% lifespan increase for the core node in the Proximity Swarm and a 1.71\% increase for the downstream outlet node in the Monolithic cluster. This tangible extension of physical durability is the primary mathematical pathway to achieving sustainable orbital AI.

\textbf{Throughput Stabilization as a Secondary Byproduct}. Evaluated strictly through a terrestrial performance lens, the MFU gains, particularly the marginal 0.2\% increase in the Swarm configuration (from 90.0\% to 90.2\%), may appear susceptible to system overhead. However, this marginality validates our thesis: the Swarm does not fail due to a lack of aggregate computational capacity, but rather due to structural latency variance caused by localized thermal throttling.

Figure~\ref{fig:tail_latency} demonstrates this straggler effect. While the Baseline Monolithic architecture exhibits severe step-like latency variance capping the MFU at 75.1\% (which TLB decisively improves to 82.7\%), the absolute MFU of the Swarm was inherently high. TLB's true contribution in the Swarm is perfectly equalizing the computation time across all 36 nodes, indicated by the entirely flat TLB latency distribution. The 0.2\% MFU gain is merely a secondary mathematical byproduct of eliminating synchronization wait times. The primary objective of averting hardware destruction via thermal load balancing has been successfully achieved.

\section{Discussion and Conclusion}
Terrestrial AI's voracious appetite for energy and water is driving the industry toward ODCs. However, this transition shifts the root cause of unsustainability: while operations become strictly zero-carbon, the environmental burden is entirely front-loaded into the massive \textit{embodied carbon} of rocket launches and the impending accumulation of space e-waste. This paper highlights that the sub-$10\mu\text{s}$ latency demands of synchronized workloads force ODCs into extreme physical density, triggering a Proximity-Thermal Paradox that causes premature hardware failure and collapses MFU.

By introducing the Thermal-Aware Heterogeneity Thesis and the TLB framework, we demonstrated that resolving spatial heat traps through asymmetric software scheduling can restore synchronous throughput and mitigate cyclic thermal fatigue. Extending hardware lifespans via software orchestration is not merely a performance optimization; it is the indispensable prerequisite for amortizing the embodied carbon of space-based AI. As ODCs evolve into generalized orbital supercomputers, Thermal-Aware Heterogeneity becomes a foundational concept for any distributed scientific workload constrained by barrier synchronization, extending the relevance of TLB beyond generative AI.

To fully realize sustainable orbital computing, the sustainable computing community must confront several critical open challenges:

\textbf{Dynamic Network-Thermal Co-design}. Terrestrial AI clusters rely on dynamic network reconfiguration (e.g., Optical Circuit Switches \cite{JupiterEvolving_2022}) to optimize all-to-all communication. In space, reconfiguring FSO links within a proximity swarm requires mechanical or optical steering. This directly alters a node's physical orientation, instantly changing its radiative view factors and thermal state. Co-optimizing dynamic network topologies with instantaneous orbital thermodynamics remains an entirely unsolved challenge.

\textbf{The E-Waste vs. Carbon Trade-off}. While software schedulers like TLB can extend hardware survival, orbital chips currently cannot be repaired. The community must conduct rigorous lifecycle assessments to answer a provocative question: does the "zero operational carbon" benefit of ODCs mathematically justify a paradigm of "disposable" high-end AI accelerators? Addressing this may eventually necessitate designing orbital hardware explicitly for in-space recycling.

\textbf{Federated Thermal Telemetry for Heterogeneous Swarms}. Future GW-scale ODCs will likely be multi-tenant constellations composed of hardware from diverse vendors. Executing cluster-wide thermal scheduling requires nodes to share their physical statuses. Developing open standards to broadcast "thermal margins" without exposing proprietary silicon thermal layouts or payload capabilities is essential to building cooperative, sustainable orbital ecosystems.

\section*{Acknowledgements}\label{sec:ack}
This work was supported by the Research Institute of Trustworthy Autonomous Systems (RITAS), Shenzhen Science and Technology Program (No. GJHZ20210705141807022), and Guangdong Province Innovative and Entrepreneurial Team Programme (No. 2017ZT07X386). Georgios Theodoropoulos is the corresponding author.

\bibliographystyle{ACM-Reference-Format}
\bibliography{ref}

\end{document}